\documentclass{article}
\usepackage{amsfonts}

\usepackage{graphicx}
\usepackage{amsmath}

\input{tcilatex}

\begin{document}

\title{Non-Abelian Born-Infeld action, geometry and supersymmetry.}
\author{Diego Julio Cirilo-Lombardo \\
Bogoliubov Laboratory of Theoretical Physics\\
Joint Institute for Nuclear Research, 141980, Dubna\\
Russian Federation}
\maketitle

\begin{abstract}
In this work, we propose a new non-abelian generalization of the Born-Infeld
lagrangian. It is based on a geometrical property of the abelian Born-Infeld
lagrangian in its determinantal form. Our goal is to extend the abelian
second type Born-Infeld action to the non-abelian form preserving this
geometrical property, that permits to compute the generalized volume element
as a linear combination of the components of metric and the Yang-Mills
energy-momentum tensors. Under BPS-like condition, the action proposed
reduces to that of Yang-Mills theory, independently of the gauge group. New
instanton-wormhole solution and static and spherically symmetric solution in
curved space-time for a SU(2) isotopic ansatz is solved and the N=1
supersymmetric extension of the model is performed.
\end{abstract}

\section{\protect\bigskip Introduction}

In 1934 M. Born and L. Infeld [1,3] introduced the most relevant version of
the non-linear electrodynamics with, among others, these main properties:

\textit{\ i) Geometrically} the BI Lagrangian density is one of the most
simplest non-polinomial Lagrangian densities that is invariant under the
general coordinate transformations.

ii) The BI electrodynamics is the only \textit{causal} spin-1 theory [6,8]
aside the Maxwell theory. The vacuum is characterized with $F_{\mu \nu }=0$
and the \textit{energy density} is definite semi-positive.

iii) The BI theory \textit{conserves helicity }[7] and solves the problem of
the \textit{self-energy} of the charged particles [1, 3, 23].

Recently, interest has been rising in this non-linear electromagnetic theory
since it has turned out to play an important role in the development of the
string theory, as was wery well described in the pioneering work of \
Barbashov and Chernikov [5]. The non-linear electrodynamics of the
Born-Infeld lagrangian, shown in [24], describes the low energy process on
D-branes which are non-perturbative solitonic objects that arises for the
natural D-dimensional extension of the string theory. The structure of the
string theory was improved significatively with the introduction of the
D-branes, because many physically realistic models can be constructed. For
example, the well known ''brane-world'' scenario that naturally introduces
the BI electrodynamics into the gauge theories. From the point of view of
gravity and supergravity theories, the precise form of Born-Infeld
electrodynamics on the D-brane in arbitrary background is not yet known with
certainty, principally in the case of SU(N) gauge fields [25].

With the recent advent of the physics of D-branes, the solitons in the
non-perturbative spectrum of string theory, it has been realized that their
low energy-dynamics can be properly described by the so called
Dirac-Born-Infeld (DBI) action [38, 39]. Since single branes are known to be
described by the abelian DBI action, one migth expect naturally that
multiple brane configurations would be a non abelian generalization of the
Born-Infeld action. Specifically in the case of superstring theory one has
to deal with a supersymmetric extension of DBI actions and when the number
of D-branes coincides there is a symmetry enhancement [40] and the abelian
DBI\ action should be generalized to its non abelian counterpart. Several
possibilities for extending the abelian BI\ action to the case of
non-abelian gauge symmetry have been discussed in the literature (e.g.
ref.[11]). Basically, as the starting point of all these attempts to pass to
the non abelian case of the BI action is the abelian BI\ action in its
standard form, all they differ in the way that the group trace operation is
defined. In the superstring-branes context the basic requirements that any
candidate for the NBI action will be fulfil one can mention:

i) it will not contains odd powers of the $F$ field strengh (with this
requirement one can make contact with the tree level of the open superstring
action);

ii) the action will linearize by the BPS conditions and to equations of
motion which coincides with those arising by imposing the vanishing of the $%
\beta $-function for background fields in the open superstring theory [25];

iii) if the action is linearized under BPS conditions it should be connected
with the possibility of supersymmetrizing the Born-Infeld theory.

With these significant reasons, it is interesting to generalize the
Born-Infeld lagrangian towards non-abelian electromagnetic fields.

In this work a new non-abelian generalization of the Born-Infeld action is
presented . This new non-abelian Born-Infeld action fulfil the requirements
given above and, for instance, is an admissible strong candidate for
effective action for superstrings and D-branes, besides the fundamental
importance of such non-abelian generalization has in the context of
gravitation theory and nonlinear electrodynamics . It is based in a
geometrical property of the abelian Born-Infeld lagrangian in its
determinantal form. We extend the abelian Born-Infeld action, in a similar
form, as was suggested by Hull et al. in [15], to its non-abelian
counterpart naturally preserving this geometrical identity. This fact
permits to compute the generalized volume element of the action as a linear
combination of the components of metric and the Yang-Mills energy-momentum
tensors. We show that the lagrangian proposed as a candidate for the
non-abelian Born-Infeld theory gives a very wide spectrum of gravitational
exact solutions, and also in the case of flat $O\left( 4\right) $
configurations the structure of the proposed action fulfil the energy-BPS
considerations and the topological bound given by the Minkowski's inequality
[27]: the action proposed reduces automatically to the Yang-Mills form under
BPS-like conditions. This means that the Tseytlin prescription of
symmetrized trace clearly is not the only one that takes the linear form
under BPS considerations [28]. The new non-abelian generalization of the
Born-Infeld lagrangian presented here is consistent not only of the BPS
point of view, but also from the first principles: independence of the gauge
group and conservation of its structure in all types of configurations. The
plan of this paper is as follows: in Section 2 we describe from a
geometrical point of view the non-abelian Born-Infeld (NBI) action. In
Section 3 the determinant of the NBI action is computed and the minimum
requeriments for the NBI lagrangian are enumerated. Sections 4, 5 and 6
explicitly are devoted to analyse the structure of the energy-momentum
tensor from topological considerations, the comparison with other
prescriptions and the conditions under the NBI-action\ is simplified. In
Sections 7 and 8 the dynamical equations derived from the new lagrangian
proposed and static spherically symmetric solution in curved space-time for
a SU(2) isotopic ansatz and an euclidean SU(2) instanton-wormhole are
solved. In Section 9 the $N=1$ supersymmetric extension of the non-abelian
Born-Infeld action proposed is sucessfully performed, and finally remarks
and conclusions in Section 10.

Our convention is as in ref.[2] with signatures of the metric, Riemann and
Einstein tensors all positives (+++), the internal indexes (gauge group) are
denoted by $a,b,c...$, space-time indexes by Greek letters $\mu ,\nu ,\rho
...$ and the tetrad indexes by capital latin letters $A,B,C.....$

\section{Geometrical identity and natural non-abelian generalization of the
Born-Infeld action}

As was shown in [15, 27], if $(\mathcal{M},g_{\mu \nu })$ is a riemmanian
4-manifold, then the action of the BI theory in this manifold is 
\begin{equation}
S_{BI}=\int \frac{b^{2}}{4\pi }\left( \sqrt{-g}-\sqrt{det(g_{\mu \nu }+%
\mathcal{F}_{\mu \nu })}\right) dx^{4}
\end{equation}
where we defined $g\equiv detg_{\mu \nu }$ and $\mathcal{F}_{\mu \nu
}=F_{\mu \nu }+B_{\mu \nu }$. As in [27], $B_{\mu \nu }$ is a background two
form which is not necessarily constant but: $d\mathcal{F}=0$ . Note that the
antisymmetric property of the indexes of the tensors, we have 
\begin{equation*}
det\left( g_{\mu \nu }+\mathcal{F}_{\mu \nu }\right) =det(g_{\mu \nu }-%
\mathcal{F}_{\mu \nu }) 
\end{equation*}
\begin{equation*}
\left[ det\left( g_{\mu \nu }+\mathcal{F}_{\mu \nu }\right) \right] ^{2}=g\
det(g_{\mu \nu }+\mathcal{F}_{\mu \lambda }\mathcal{F}_{\nu }^{\text{ }%
\lambda }) 
\end{equation*}
thus we can write the eq. (1) in the following form 
\begin{equation}
S_{BI}=\int \frac{b^{2}}{4\pi }\left( \sqrt{-g}-\sqrt[4]{\left| g\right| \
det\left( g_{\mu \nu }+\mathcal{F}_{\mu \lambda }\mathcal{F}_{\nu }^{\
\lambda }\right) }\right) dx^{4}
\end{equation}
From this form of the determinant, the natural non-abelian generalization of
the Born-Infeld action is 
\begin{equation}
S_{NBI}=\int \frac{b^{2}}{4\pi }\left( \sqrt{-g}-\sqrt[4]{\left| g\right| \
det\left( g_{\mu \nu }+\mathcal{F}_{\ \mu \lambda }^{a}\mathcal{F}_{a\nu
}^{\ \ \ \lambda }\right) }\right) dx^{4}
\end{equation}
From the expansion of the determinant in eq.(3) (we turn $B_{\mu \nu }$=0
for simplicity) 
\begin{equation}
S_{NBI}=\frac{b^{2}}{4\pi }\int \sqrt{-g}dx^{4}\left\{ 1-\sqrt[4]{\gamma
^{4}-\frac{\gamma ^{2}}{2}\overline{M}^{2}-\frac{\gamma }{3}\overline{M}^{3}+%
\frac{1}{8}\left( \overline{M}^{2}\right) ^{2}-\frac{1}{4}\overline{M}^{4}}%
\right\}  \tag{(3a)}
\end{equation}
where we defined 
\begin{equation}
\begin{array}{lllll}
M_{\mu \nu }\equiv F_{\ \mu \lambda }^{a}F_{a\nu }^{\ \ \ \lambda } & ; & 
\gamma \equiv \left( 1+\frac{F_{\ \mu \lambda }^{a}F_{a}^{\ \ \ \mu \lambda }%
}{4}\right) & ; & \overline{M}_{\mu \nu }\equiv M_{\mu \nu }-\frac{g_{\mu
\nu }}{4}F_{\ \alpha \beta }^{a}F_{a}^{\alpha \beta } \\ 
&  &  &  &  \\ 
\overline{M}_{\rho }^{\nu }\overline{M}_{\nu }^{\rho }\equiv \overline{M}^{2}
& ; & \overline{M}_{\lambda }^{\nu }\overline{M}_{\rho }^{\lambda }\overline{%
M}_{\nu }^{\rho }\equiv \overline{M}^{3} & ; & \left( \overline{M}_{\rho
}^{\nu }\overline{M}_{\nu }^{\rho }\right) ^{2}\equiv \left( \overline{M}%
^{2}\right) ^{2} \\ 
&  &  &  &  \\ 
\overline{M}_{\mu }^{\nu }\overline{M}_{\lambda }^{\mu }\overline{M}_{\rho
}^{\lambda }\overline{M}_{\nu }^{\rho }\equiv \overline{M}^{4} &  &  &  & 
\\ 
&  &  &  & 
\end{array}
\tag{(3b)}
\end{equation}
and notice that all quantities into the fourth root in $S_{NBI}$ are
adimensionalized over the Born-Infeld absolute field $b$. We can see that
only one general invariant $S\equiv -\frac{1}{4}F_{\mu \nu }^{a}F_{a}^{\mu
\nu }$ of the electromagnetic field is the basic block for the exension of
the Born-Infeld lagrangian to its non-abelian counterpart (not particular
trace prescriptions). The pseudoscalar invariant $P\equiv -\frac{1}{4}F_{\mu
\nu }\widetilde{F}^{\mu \nu }$ is restored only when we pass to the $U\left(
1\right) $ gauge group; in contrast with the non-abelian generalization
(dependent on the gauge group) of the Born-Infeld action proposed by
Hagiwara [4], that has three invariants of the nine possible gauge
invariants for SU$\left( 2\right) $Yang-Mills fields [10].

In resume, from the abelian Born-Infeld action (2) as the starting point,
its non-abelian generalization is performed in a natural form (3) and we can
see that: \textit{first}, in the NBI action proposed only $F_{\mu \nu
}^{a}F_{a}^{\mu \nu }$ is the basic block for its construction; and \textit{%
second}, the full form of the action (3) it is independent of the gauge
group.

\section{Requirements for the non-abelian generalization and explicit
computation of the determinant}

By analogy with the abelian case, the lagrangian will satisfy the following
properties [16]

i) We should find the ordinary Yang-Mills theory in the limit $b\rightarrow
\infty .$

ii) the electric components $F_{\mu 0}^{a}$ \footnote{%
Here the zero index corresponds to the time.} non-abelian electromagnetic
tensor should be bounded for i) when the magnetic components vanish 
\footnote{%
The polynomial under the root should start with terms as 1-$\frac{\left(
F_{0\mu }^{a}\right) ^{2}}{b^{2}}+...$ when the magnetic components are zero.%
}

iii) The action will be invariant under diffeomorphisms .

iv) The action will be real.

The proposed non-abelian Born-Infeld action (3) fulfil all the propierties
described above, as we will see in the explicit computation of the
determinant . For tensors represented by matrices or hypermatrices, the
algebraic invariant associated to a matrix \textbf{a }can be obtained as
traces of the powers of the given matrix. According to the Cayley-Hamilton
theorem [20,21,22], only a finite number of these powers is linearly
independent and therefore only a finite number of invariants is linearly
independent. A more convenient set of invariants is given by the
discriminants which are suitable combinations of traces and constructed in
terms of alternating products with the unit matrix $\mathbb{I}$.

In order to perform the explicit computation of the determinant, we start
defining the following 
\begin{equation}
\left\langle \mathbf{a}\right\rangle \equiv tr\left( \mathbf{a}\right) =%
\mathbf{a}_{i}^{i}\ \ \ \ \ \ \ \ \ \ \ \ \ \ \ \ \ \left\langle \mathbf{a}%
^{s}\right\rangle \equiv tr\left( \mathbf{a}^{s}\right) =\left( \mathbf{a}%
^{s}\right) _{i}^{i}
\end{equation}
The relation between the traces of different powers of \ $\mathbf{a}$ and
its determinant in D=4 is 
\begin{equation}
det(\mathbf{a)}=\frac{1}{4!}\left[ \ \left\langle \mathbf{a}\right\rangle
^{4}-6\left\langle \mathbf{a}\right\rangle ^{2}\left\langle \mathbf{a}%
^{2}\right\rangle +8\left\langle \mathbf{a}\right\rangle \left\langle 
\mathbf{a}^{3}\right\rangle +3\left\langle \mathbf{a}^{2}\right\rangle
^{2}-6\left\langle \mathbf{a}^{4}\right\rangle \right]
\end{equation}

If $\mathbf{a=}\left( g_{\mu \nu }+F_{\mu \lambda }^{a}F_{\nu a}^{\ \lambda
}\right) $, the second factor into the square root in (3) takes the familiar
form 
\begin{equation}
det\left( g_{\mu \nu }+F_{\mu \lambda }^{a}F_{\nu a}^{\ \lambda }\right) =%
\text{ \ \ \ \ }
\end{equation}
\begin{equation*}
=g\left[ \gamma ^{4}-\frac{\gamma ^{2}}{2}\overline{M}^{2}-\frac{\gamma }{3}%
\overline{M}^{3}+\frac{1}{8}\left( \overline{M}^{2}\right) ^{2}-\frac{1}{4}%
\overline{M}^{4}\right] 
\end{equation*}
now we can easily seen that the proposed NBI action (3)\ fulfil the four
requirements given above.

\section{Energy-momentum tensor}

From the determinant form of the NBI lagrangian we can obtain the energy
momentum tensor, varying eq.(3) with respect to $g_{\mu \nu }$ as usual 
\begin{equation}
-\frac{1}{2}T_{\mu \nu }\equiv \frac{1}{\sqrt{g}}\frac{\delta \mathcal{L}}{%
\delta g^{\mu \nu }}=\frac{1}{\sqrt{g}}\frac{\delta \left( \sqrt{g}%
L_{NBI}\right) }{\delta g^{\mu \nu }}
\end{equation}
we obtain in matrix form 
\begin{equation}
T_{\mu \nu }=\frac{b^{2}}{4\pi }\left\{ g_{\mu \nu }-\frac{\sqrt[4]{g+M}}{2\ 
\sqrt[4]{g}}\left[ g_{\mu \nu }-\left( g+M\right) _{\sigma \rho }^{-1}\left(
F_{\mu \sigma }^{a}F_{\nu \rho a}^{\ }-g_{\mu \sigma }g_{\nu \rho }\right) %
\right] \right\}
\end{equation}
another useful quantity that is the trace of the energy-momentum tensor 
\begin{equation}
T_{\mu }^{\mu }=\frac{b^{2}}{4\pi }\left\{ g_{\mu }^{\mu }-\frac{\sqrt[4]{g+M%
}}{2\ \sqrt[4]{g}}\left[ g_{\mu }^{\mu }-\left( g+M\right) _{\sigma \rho
}^{-1}\left( F_{\mu \sigma }^{a}F_{\rho a}^{\mu \ }-g_{\sigma \rho }\right) %
\right] \right\}
\end{equation}
From the expression for the energy momentum tensor, we can obtain under
which conditions it nullifies i.e: $T_{\mu \nu }=0$%
\begin{equation}
F_{\mu \lambda }^{a}F_{\nu a}^{\ \lambda }=g_{\mu \nu }\frac{F_{\rho \lambda
}^{a}F_{a}^{\ \rho \lambda }}{4}
\end{equation}
This shows that in the combined system 
\begin{equation}
I=I\left( F\right) +\int_{M}d^{4}x\sqrt{g}R\left( g\right)
\end{equation}
this action is extremized on an Einstein manifold with an (anti) self-dual
field configuation. Notice also, that expression (10) is the condition under
which the self dual configuration saturates the Minkowski's inequality for
arbitrary gauge group [27].

Some remarks: i) it was shown in reference [27] that in a general curved
background metric the Born-Infeld action is bounded by a topological
quantity, and the bound is realized when the gauge field configuration is
(anti-)self-dual. This means that from pure topological considerations the
non-abelian Born-Infeld action must satisfy this condition.

ii) by construction our lagrangian satisfies the above condition
automatically, saturating the Minkowski's inequality (topological bound)
when the (anti-)self-duality condition (10) is inserted in (3). In this case
the NBI\ lagrangian linearizes satisfying also the BPS-like conditions [28],
becoming the Yang-Mills lagrangian.

iii) from the point ii) we can see that our action is minimal remained
closely (below) the topological bound more than the other non-abelian
lagrangians proposed.

These remarks becomes important when one studies instanton configurations in
any background metric. The fact that any instanton configuration does not
affect the background metric can be seen clearly in the context of the
superstring theory. In equations of motion of supergravity, under a
particular ansatz of the dilaton and R-R scalar field concerning self-dual
point, the energy momentum tensor of them vanishes[34,35]. For instance,
these D-instantons do not affect the Einstein equation. The inclusion of the 
$B_{\mu \nu }$ was disscussed in detail in ref.[36].

\section{Comparison with other prescriptions}

Following the notation in ref.[16], we have the folowing expression relating
the determinant of a linear operator $A$ to traces 
\begin{equation*}
\begin{array}{l}
\left[ \det \left( 1+A\right) \right] ^{\beta }=\exp \left\{ \beta tr\left[
\log \left( 1+A\right) \right] \right\} \\ 
\\ 
=\overset{\infty }{\underset{n=0}{\sum }}\;\underset{\underset{\in \left[
S_{n}\right] }{\underset{-}{\alpha }=\left( \alpha _{1}....\alpha
_{n}\right) }}{\sum }\left( -1\right) ^{n}\overset{n}{\underset{p=1}{\prod }}%
\frac{1}{\alpha _{p}!}\left( -\frac{\beta tr\left( A^{p}\right) }{p}\right)
^{\alpha _{p}}%
\end{array}
\end{equation*}
where $\underset{-}{\alpha }\in \left[ S_{n}\right] $ and $\left[ S_{n}%
\right] $ is the set of equivalence classes of the permutation group of
order $n$. The multi-index $\underset{-}{\alpha }$ is given by a
Ferrer-Young diagram that equivalently satisfies the following relation 
\begin{equation*}
\overset{n}{\underset{p=1}{\sum }}p\alpha _{p}=n,\;\;\;\;\alpha _{p}\geq 0 
\end{equation*}
In order to analyze the structure of different lagrangians up to any order
in $F$ we can use this trace formula. We denote : $tr_{R}$ $\equiv $ trace
over representation indexes, $tr_{\otimes }\equiv $ trace over the tensor
product and $tr_{\mathcal{M}}\equiv $ the trace taken over space-time
indexes. For our proposed lagrangian (3) we have 
\begin{equation*}
\begin{array}{l}
\left[ \det_{\mathcal{M}}\left( 1+M\right) \right] ^{1/4}=\overset{\infty }{%
\underset{n=0}{\sum }}\;\underset{\underset{-}{\alpha }=\left( \alpha
_{1}....\alpha _{n}\right) }{\sum }\left( -1\right) ^{n}\overset{n}{\underset%
{k=1}{\prod }}\frac{1}{\alpha _{k}!}\left( -\frac{1}{4}\frac{tr_{\mathcal{M}%
}\left( M^{k}\right) }{k}\right) ^{\alpha _{k}} \\ 
\\ 
=\overset{\infty }{\underset{n=0}{\sum }}\;\underset{\underset{-}{\alpha }%
=\left( \alpha _{1}....\alpha _{n}\right) }{\sum }\left( -1\right) ^{n}%
\overset{n}{\underset{k=1}{\prod }}\frac{1}{\alpha _{k}!}\overset{k}{%
\underset{\alpha _{k}\neq 0}{\underset{m=1}{\prod }}}\left( -\frac{1}{4k}tr_{%
\mathcal{M}}M^{m}\right) \ \hspace{1cm}\hspace{1cm};M_{\alpha \gamma }\equiv
F_{\alpha \beta }^{a}F_{a\gamma }^{\ \ \beta }%
\end{array}
\end{equation*}
The expansion for the symmetrized trace prescription given by Tseytlin in
[11] is given by 
\begin{equation*}
\begin{array}{l}
\frac{1}{d_{R}}Str_{R}\left[ \det_{\mathcal{M}}\left( 1+iF^{a}T_{a}\right) %
\right] ^{1/2}= \\ 
\\ 
=\frac{1}{d_{R}}Str_{R}\overset{\infty }{\underset{n=0}{\sum }}\;\underset{%
\underset{-}{\alpha }=\left( \alpha _{1}....\alpha _{n}\right) }{\sum }%
\left( -1\right) ^{n}\overset{n}{\underset{k=1}{\prod }}\frac{1}{\alpha _{k}!%
}\left( -\frac{tr_{\mathcal{M}}\left( F^{a_{1}}......F^{a_{2k}}\right) }{4k}%
T_{a_{1}}....T_{a_{2k}}\right) ^{\alpha _{k}} \\ 
\\ 
=\overset{\infty }{\underset{n=0}{\sum }}\;\underset{\underset{-}{\alpha }%
=\left( \alpha _{1}....\alpha _{n}\right) }{\sum }\left( -1\right) ^{n}\left[
\overset{n}{\underset{k=1}{\prod }}\frac{1}{\alpha _{k}!}\overset{k}{%
\underset{\alpha _{k}\neq 0}{\underset{m=1}{\prod }}}\left( -\frac{tr_{%
\mathcal{M}}\left( F^{a_{1}^{m}}......F^{a_{2k}^{m}}\right) }{4k}\right) 
\frac{1}{d_{R}}Str_{R}\left( \overset{n}{\underset{k=1}{\prod }}\overset{n}{%
\underset{m=1}{\prod }}T_{a_{1}^{m}}....T_{a_{2k}^{m}}\right) \right]%
\end{array}
\end{equation*}
and for the trace prescription inspired by non-commutative geometry given by
E. Serie' et al. in [16] we have 
\begin{equation*}
\begin{array}{l}
\left[ \det_{\mathcal{M}\otimes R}\left( 1+F^{2}\right) \right] ^{1/4d_{R}}=%
\overset{\infty }{\underset{n=0}{\sum }}\;\underset{\underset{-}{\alpha }%
=\left( \alpha _{1}....\alpha _{n}\right) }{\sum }\left( -1\right) ^{n}%
\overset{n}{\underset{k=1}{\prod }}\frac{1}{\alpha _{k}!}\left( -\frac{tr_{%
\mathcal{\otimes }}\left( F^{2k}\right) }{d_{R\times }k}\right) ^{\alpha
_{k}} \\ 
\\ 
=\overset{\infty }{\underset{n=0}{\sum }}\;\underset{\underset{-}{\alpha }%
=\left( \alpha _{1}....\alpha _{n}\right) }{\sum }\left( -1\right) ^{n}%
\overset{n}{\underset{k=1}{\prod }}\frac{1}{\alpha _{k}!}\overset{k}{%
\underset{\alpha _{k}\neq 0}{\underset{m=1}{\prod }}}\left( -\frac{tr_{%
\mathcal{M}}\left( F^{a_{1}^{m}}......F^{a_{2k}^{m}}\right) }{4k}\right)
tr_{R}\left( \frac{T_{a_{1}^{m}}....T_{a_{2k}^{m}}}{d_{R}}\right)%
\end{array}
\end{equation*}
It is very easy to see from the above expansions the following:

i) In the three cases, the third-order and higher order invariants of the
non-abelian electromagnetic field do not appear.

ii) Clearly Serie' et al. and Tseytlin prescriptions have a high dependence
on the gauge group, that is not in our proposed non-abelian lagrangian
because the basic block of the lagrangian (3) is $M_{\alpha \gamma }\equiv
F_{\alpha \beta }^{a}F_{a\gamma }^{\ \ \ \beta }$.

In this sense, the NBI lagrangian (3) has its gauge structure ''hidden''
into the basic object $M$ .This is the reason why the the lagrangian (3)
remains closely by below to the topological bound given by the Minkowski's
inequality more than the other non-abelian proposals. And this is important
from the BPS considerations and their relation with field configurations
with minimal energy as in worldvolume theories of D\textit{p}-branes with
specific values of \textit{p} admitting '' worldvolume (BPS) solitons''
[28]. It should be mentioned that the differences between our action and the
symmetrized trace prescription, that for the usual gauge groups and for the
leading terms in the expansions given above have not relevant importance in
the context of the string theory, they can turn very important in the brane
theories. In particular, there are some discrepancies between the results
arising from a symmetrized non-abelian Born-Infeld theory and the spectrum
to be expected in brane theories as pointed out in ref.[37]. This fact will
be tested with our proposed NABI\ generalization in the same brane context
in order to see whether there exist similar discrepancies or not in future
work [26]. It must be noted that this discussion should be viewed from
within the context of a general analysis of possible NABI\ actions.

\section{Topology of the gauge fields and space-time: The reduced lagrangian}

Despite the aparent complexity of the non abelian lagrangian (3), there
exists a possibility in which it can be reduced to a square root form. This
possibility is related with the choice of an ansatz where the gauge group
and space-time are highly identified (high symmetry). The requirement for
that expression into the fourth root of the action (3) becomes a perfect
square is given when the following factorization propierty for the traceless 
$\overline{M}$ holds 
\begin{equation}
\overline{M}_{\alpha \beta }\overline{M}^{\gamma \beta }\approx \delta
_{\alpha }^{\gamma }\mathbb{Y}\left( F_{\mu \nu }^{a}F_{a}^{\mu \nu }\right)
\end{equation}
where the scalar function $\mathbb{Y}$ depends obviously on the invariant $%
F_{\mu \nu }^{a}F_{a}^{\mu \nu }.$ Several well known ansatz reduce the form
of the fourth root lagrangian (3) to square root form (e.g. Ogura and Hosoya
[30], etc) 
\begin{equation*}
S_{NBI}=\frac{b^{2}}{4\pi }\int \sqrt{-g}\left( 1-\mathbb{R}\right) dx^{4} 
\end{equation*}
where now 
\begin{equation}
\mathbb{R\equiv }\sqrt{\ 1+\frac{1}{2b^{2}}\left( F_{\ \rho \lambda
}^{a}F_{a}^{\ \rho \lambda }\right) -\frac{1}{4b^{4}}\left( F_{\ \lambda
a}^{\nu }F_{\rho }^{a\ \lambda }\right) \left( F_{\ \lambda a}^{\rho }F_{\nu
}^{a\ \lambda }\right) +\frac{1}{8b^{4}}\left( F_{\ \rho \lambda
}^{a}F_{a}^{\ \rho \lambda }\right) ^{2}}
\end{equation}

Notice that when the requirement (12) holds \textit{only} the traces of even
products of $\overline{M}$'s they will appear in the explicit computation of
the determinant into the root of (3) given the following result

\begin{equation}
det\left( g_{\mu \nu }+\overline{F}_{\mu \lambda }\overline{F}_{\nu }^{\
\lambda }\right) =\text{ \ \ \ \ }
\end{equation}
\begin{equation*}
=g^{2}\left[ 1+\frac{1}{2b^{2}}\left( F_{\ \rho \lambda }^{a}F_{a}^{\ \rho
\lambda }\right) -\frac{1}{4b^{4}}\left( F_{\ \lambda a}^{\nu }F_{\rho }^{a\
\lambda }\right) \left( F_{\ \lambda a}^{\rho }F_{\nu }^{a\ \lambda }\right)
+\frac{1}{8b^{4}}\left( F_{\ \rho \lambda }^{a}F_{a}^{\ \rho \lambda
}\right) ^{2}\right] ^{2} 
\end{equation*}
compare with expression (6). This property will be used when we solve and
analyze different configurations with non-abelian fields in the following
sections.

\section{Equations of motion for the non-abelian Born-Infeld theory in
curved space-time}

Now, we pass to describe the dynamical equations for the non-abelian
electromagnetic fields in curved space-time. Geometrically, introducing the
generalizated exterior derivative $d$, the equations can be written 
\begin{equation}
dF=0\ \ \ \ \ \ ;\ \ \ \ \ \ d\widetilde{\mathbb{F}}=0
\end{equation}
where (15) in components and in a orthonormal frame (tetrad) are 
\begin{equation*}
D_{B}\widetilde{F}^{ABa}=0 
\end{equation*}
\begin{equation}
D_{B}\mathbb{F}^{ABa}=\nabla _{B}\mathbb{F}^{ABa}+\varepsilon ^{abc}A_{B}^{a}%
\mathbb{F}^{ABc}
\end{equation}
where 
\begin{equation}
\left( \nabla _{B}\mathbb{F}^{AB}\right) ^{a}\equiv \frac{\delta \mathcal{L}%
}{\delta F_{ABa}}=\left( e_{B}^{\ \ \mu }\ \mathbb{F}_{,\mu }^{AB}+\Gamma
_{CD}^{A}\mathbb{F}^{CD}+\Gamma _{BD}^{D}\mathbb{F}^{AB}\right) ^{a}
\end{equation}
and 
\begin{equation}
\mathbb{F}_{\ bc}^{a}\equiv \frac{\partial L_{NBI}}{\partial F_{a}^{\ bc}}
\end{equation}
The Bianchi identity is automatically satisfied because, obviously: $%
F^{a}=dA^{a}+A^{b}\wedge A^{c}$ . $\mathbb{F}^{ABa}$ is the non-abelian
extension of the Born-Infeld dielectric displacement-like tensor in the
abelian Born-Infeld electrodynamics.

Trivial solutions to the dynamical equations of the non-abelian
electromagnetic fields are easily obtained, from the action (3). We will
solve one of these cases as an example.

\paragraph{Spherically symmetric chromostatic solution}

The equations that describe the dynamic of the non-abelian electromagnetic
fields of the Born-Infeld theory in a curved space-time are (16).
Explicitly, and for the\textit{\ isotopic gauge }[9], are 
\begin{equation}
dF^{a}=d\left( F_{01}^{a}\omega ^{0}\wedge \omega ^{1}\right) =\partial
_{\theta }\left( e^{\Lambda +\Phi }F_{01}^{a}\right) d\theta \wedge dt\wedge
dr=0
\end{equation}
\begin{equation}
d\mathbb{F}^{a}=d\left( -\mathbb{F}_{01}^{a}\omega ^{3}\wedge \omega
^{2}\right) =\partial _{r}\left( -e^{2G}\mathbb{F}_{01}^{a}\right) dr\wedge
d\varphi \wedge d\theta =0
\end{equation}
where the 1-forms $\omega ^{\alpha }$ correspond to the spherically
symmetric interval 
\begin{equation}
ds^{2}=-e^{2\Lambda }dt^{2}+e^{2\Phi }dr^{2}+e^{2F\left( r\right) }\left(
d\theta ^{2}+\sin ^{2}\theta \,\right) d\varphi ^{2}
\end{equation}
Explicitly 
\begin{equation}
\begin{array}{cccc}
\omega ^{0}=e^{\Lambda }dt &  & \Rightarrow & dt=e^{-\Lambda }\omega ^{0} \\ 
\omega ^{1}=e^{\Phi }dr &  & \Rightarrow & dr=e^{-\Phi }\omega ^{1} \\ 
\omega ^{2}=e^{F\left( r\right) }d\theta &  & \Rightarrow & d\theta
=e^{-F\left( r\right) }\omega ^{2} \\ 
\omega ^{3}=e^{F\left( r\right) }\sin \theta \,d\varphi &  & \Rightarrow & 
d\varphi =e^{-F\left( r\right) }\left( \sin \theta \right) ^{-1}\omega ^{3}%
\end{array}%
\end{equation}
From (19,20) we can see that 
\begin{equation}
F_{01}^{a}=f\left( r,a\right)
\end{equation}
and 
\begin{equation}
e^{2G}\mathbb{F}_{01}^{a}=h\left( a\right)
\end{equation}
The simplest form for the fields and the potential that belongs from the
above expression is 
\begin{equation}
F_{01}^{a}=f\left( r,a\right) =-\delta ^{az}\nabla _{r}\phi \left( r\right)
\ \ \ \ \ \ \ \text{and}\ \ \ \ \ \ \ \ e^{2G}\mathbb{F}_{01}^{a}=h\left(
a\right) =const\ \ 
\end{equation}
On the other hand, from (18) we can see that 
\begin{equation}
\mathbb{F}_{01}^{a}=\frac{F_{01}^{a}}{\sqrt{1-\left( \overline{F}%
_{01}\right) ^{2}}}\ \ \ \ \ \ \ \ ;\ \ \ \ \ \left( \left( \overline{F}%
_{01}\right) ^{2}\equiv \frac{F_{01}^{a}F_{a}^{01}}{b^{2}}\right)
\end{equation}
then, inverting (26)\ together with (25) 
\begin{equation}
F_{01}^{a}=\frac{b\left( a\right) }{\sqrt{1+\left( \frac{b}{h}e^{G}\right)
^{4}}}
\end{equation}
In order to obtain the final expression for the non-abelian electric field
it is necessary to introduce (27) into the Einstein equations for the
interval (21) [Appendix] (for details of full computations in the abelian
case, see [13]). Having made it, the non abelian electromagnetic field for a
static spherically symmetric configuration is 
\begin{equation}
F_{01}^{a}=\frac{b\delta ^{az}}{\sqrt{1+\left[ 1-\left( \frac{r_{0}}{\lambda
\left| r\right| }\right) ^{3}\right] ^{4}\left( \frac{r}{r_{0}}\right) ^{4}}}%
=-\delta ^{az}\nabla _{r}\phi \left( r\right)
\end{equation}
where we have associated $h=br_{0}^{2}=Q$ [1,3] and $\lambda $ is a
dimensionless parameter restricted to $\left| \lambda \right| <1$ [13] .
Integrating (28) we obtain the following expression for the potential $\phi
\left( r\right) $%
\begin{equation}
\phi \left( r\right) =-\left( -1\right) ^{1/4}br_{0}F\left[ Arcsin\left[
\left( -1\right) ^{3/4}\overline{Y}\left( r\right) ,-1\right] \right]
\end{equation}
where 
\begin{equation*}
\overline{Y}\left( r\right) \equiv \left[ 1-\left( \frac{r_{0}}{a\left|
r\right| }\right) ^{3}\right] ^{2}\frac{r}{r_{0}} 
\end{equation*}
and $F$ is the Elliptic function of the first kind [12]. However in
non-linear electrodynamics, the solution turns of Coulomb type
asymptotically ($r\rightarrow \infty $), but is strongly modified near the
origin ($r\rightarrow 0$) presenting a regular behaviour without
singularities [13]. Perhaps the isotopic ansatz is very ''knaive'' (embedded 
$U\left( 1\right) $ solution) the asymptotic behaviour of the solution is in
complete agreement with the type of solutions given by Ikeda and Miyachi [9]
for Yang-Mills in flat space.

\subparagraph{About BPS and string theory considerations}

In reference [29] the autor finds an $SO\left( 4\right) $ invariant solution
in four dimensional Euclidean space adopting the ansatz of [33] into the
following lagrangian 
\begin{equation}
S_{Park}[F,g]=\int_{R^{4}}\alpha \left( \det_{\mathcal{M}\otimes R}\left|
g_{\mu \nu }\otimes \mathbf{1}_{d_{R}}+\beta ^{-1}F_{\mu \nu }^{a}\otimes
T_{a}\right| ^{1/2d_{R}}-\sqrt{\left| g\right| }\right)
\end{equation}
obtaining, after the ansatz of [33] was introduced that this action eq. (32)
does not becomes to the Yang-Mills lagrangian under BPS condition. The main
reason why this situation happens is because the expansion of the
determinant into (32) contains odd powers of the fields and this fact, see
[28] for details, makes impossible for any proposed lagrangian
(independently of its non-linearity or complexity) to become the Yang-Mills
form, when the BPS condition is introduced.

The non-abelian lagrangian proposed in our work straightforwardly linearizes
when the ansatz of reference [33] is introduced because the determinant is
reduced to a sum of squares. Notice that this discussion should be viewed
from the context of a general analysis of possible NBI. This means that: the
action relevant to the description of multiple D-branes has its origin in
superstring theory coupled to a non-abelian gauge field; and it is well
known that the effective action of the latter does not contain a term of the
form $F^{3}$ [28]. We can see that our NBI\ action has the BPS properties
discussed and from the point of view of string theory [11] the NBI action
presented here is a strong candidate as non-abelian effective action.

\section{Wormhole-instanton solution in NABI theory}

The action is given by 
\begin{equation}
S=-\frac{1}{16\pi G}\int d^{4}x\sqrt{g}R+\int d^{4}x\sqrt{g}\Lambda +\frac{1%
}{4\pi }\int d^{4}x\sqrt{g}L_{NBI}
\end{equation}
\begin{equation*}
L_{NBI}=\left( \frac{b^{2}}{4\pi }\right) \left( 1-\mathbb{R}\right) 
\end{equation*}
\begin{equation*}
\mathbb{R\equiv }\sqrt[4]{\gamma ^{4}-\frac{\gamma ^{2}}{2}\overline{M}^{2}-%
\frac{\gamma }{3}\overline{M}^{3}+\frac{1}{8}\left( \overline{M}^{2}\right)
^{2}-\frac{1}{4}\overline{M}^{4}} 
\end{equation*}
\begin{equation*}
R^{\mu \nu }-\frac{1}{2}g^{\mu \nu }(R-\Lambda )=8\pi GT^{\mu \nu } 
\end{equation*}
Scalar curvature $R$ and th $SU\left( 2\right) $ Yang-Mills field strength $%
F_{\mu \nu }^{a}$ are defined in terms of the affine connection $\Gamma
_{\mu \nu \text{ }}^{\lambda }$ and the SU(2) gauge connection $A_{\ \mu 
\text{ }}^{a}$by 
\begin{equation}
R=g^{\mu \nu }R_{\mu \nu }\hspace{1cm}R_{\mu \nu }=R_{\mu \lambda \nu
}^{\lambda }
\end{equation}
\begin{equation*}
R_{\mu \lambda \nu }^{\lambda }=\partial _{\nu }\Gamma _{\mu \rho \text{ }%
}^{\lambda }-\partial _{\rho }\Gamma _{\mu \nu \text{ }}^{\lambda }+... 
\end{equation*}
\begin{equation*}
F_{\ \mu \nu }^{a}=\partial _{\mu }A_{\ \nu \text{ }}^{a}-\partial _{\nu
}A_{\ \mu \text{ }}^{a}+\varepsilon _{bc}^{a}A_{\ \mu \text{ }}^{b}A_{\ \nu 
\text{ }}^{c} 
\end{equation*}
$G$ and $\Lambda $ are the Newton gravitational constant and the
cosmological constant respectively. As in the case of Einstein-Yang -Mills
systems, for our non-abelian BI\ model it can be interpreted as a prototype
of gauge theories interacting with gravity (e.g. QCD, GUTs, etc). Upon
varying the action (31), we obtain the Einstein equation

\begin{equation}
R_{\mu \nu }-\frac{1}{2}g_{\mu \nu }R=8\pi G\left( T_{\mu \nu }-\Lambda
g_{\mu \nu }\right)
\end{equation}
\begin{equation*}
T_{\mu \nu }=\frac{b^{2}}{4\pi }\left\{ g_{\mu \nu }-\frac{\sqrt[4]{g+M}}{2\ 
\sqrt[4]{g}}\left[ g_{\mu \nu }-\left( g+M\right) _{\sigma \rho }^{-1}\left(
F_{\mu \sigma }^{a}F_{\nu \rho a}^{\ }-g_{\mu \sigma }g_{\nu \rho }\right) %
\right] \right\} 
\end{equation*}
and the Yang-Mills field equation in differential form 
\begin{equation}
d^{*}\mathbb{F}^{a}+\frac{1}{2}\varepsilon ^{abc}\left( A_{b}\wedge ^{*}%
\mathbb{F}_{c}-^{*}\mathbb{F}_{b}\wedge A_{c}\right) =0
\end{equation}
where we define as usual 
\begin{equation*}
\mathbb{F}_{\ bc}^{a}\equiv \frac{\partial L_{NBI}}{\partial F_{a}^{\ bc}} 
\end{equation*}
we are going to seek for a classical solution of eqs. (33) and (34) with the
following spherically symmetric ansatz for the metric and gauge connection 
\begin{equation}
ds^{2}=d\tau ^{2}+a^{2}\left( \tau \right) \sigma ^{i}\otimes \sigma
^{i}\equiv d\tau ^{2}+e^{i}\otimes e^{i}
\end{equation}
here $\tau $ is the euclidean time and the dreibein is defined by $%
e^{i}\equiv a^{2}\left( \tau \right) \sigma ^{i}.$The gauge connection is 
\begin{equation}
A^{a}\equiv A_{\mu }^{a}dx^{\mu }=h\sigma ^{a}
\end{equation}
The $\sigma ^{i}$ one-form satisfies the $SU\left( 2\right) $ Maurer-Cartan
structure equation 
\begin{equation}
d\sigma ^{a}+\varepsilon _{\ bc}^{a}\sigma ^{b}\wedge \sigma ^{c}=0
\end{equation}
Notice that in the ansatz the frame and isospin indexes are identified and
the four root NBI lagrangian (3) reduces to a square root expression, as we
explained in Section 6. The field strength two-form 
\begin{equation}
F^{\gamma }=\frac{1}{2}F_{\ \mu \nu }^{\gamma }dx^{\mu }\wedge dx^{\nu }
\end{equation}
becomes 
\begin{eqnarray}
F^{a} &=&dA^{a}+\frac{1}{2}\varepsilon _{\ bc}^{a}A^{b}\wedge A^{c} \\
&=&\left( -h+\frac{1}{2}h^{2}\right) \varepsilon _{\ bc}^{a}\sigma
^{b}\wedge \sigma ^{c}  \notag
\end{eqnarray}
Inserting it in the Yang-Mills field equation (34) and into the NBI
lagrangian we obtain 
\begin{equation}
\begin{array}{l}
d^{*}\mathbb{F}^{a}+\frac{1}{2}\varepsilon ^{abc}\left( A_{b}\wedge ^{*}%
\mathbb{F}_{c}-^{*}\mathbb{F}_{b}\wedge A_{c}\right) =0 \\ 
\\ 
=\varepsilon _{bc}^{a}d\tau \wedge \sigma ^{b}\wedge \sigma ^{c}\left(
-2h+h^{2}\right) \left( h-1\right) \left( \mathbb{A}/a\right)%
\end{array}%
\end{equation}
where 
\begin{equation*}
\mathbb{A\equiv }\frac{\left[ 1+2\left( \frac{r_{0}}{a}\right) ^{4}\left( -h+%
\frac{h^{2}}{2}\right) \right] }{\mathbb{R}} 
\end{equation*}
\begin{equation*}
^{\ast }\mathbb{F}=\mathbb{A}\left( -2h+h^{2}\right) d\tau \wedge \frac{e^{a}%
}{a^{2}} 
\end{equation*}
\begin{equation*}
\mathbb{R=}\sqrt{1+6\left( -h+\frac{h^{2}}{2}\right) ^{2}\left( \frac{r_{0}}{%
a}\right) ^{4}+6\left( -h+\frac{h^{2}}{2}\right) ^{4}\left( \frac{r_{0}}{a}%
\right) ^{8}} 
\end{equation*}

We can see that for $h=1$ there exist a non trivial solution 
\begin{equation*}
F_{bc}^{a}=-\frac{\varepsilon _{bc}^{a}}{a^{2}}\hspace{1.54cm}F_{bc}^{a}=0 
\end{equation*}
Namely, only the magnetic field is non vanishing while the electric field
vanishes. An analogous feature can be seen in the solution of Giddings and
Strominger [31]. Substituting the expression for the field strength (39)
into the Born-Infeld energy-momentum tensor, we reduce the Einstein equation
(30) to an ordinary differential equation for the scale factor $a$, 
\begin{equation}
3\left[ \left( \frac{\overset{.}{a}}{a}\right) ^{2}-\frac{1}{a^{2}}\right]
=2G\left( b^{2}-4\pi \Lambda \right) -2Gb^{2}\left[ 1+6\left( \frac{r_{0}}{a}%
\right) ^{4}\left( 1+\left( \frac{r_{0}}{a}\right) ^{4}\right) \right] ^{1/2}
\end{equation}
where the relation with the $H$ constant is given by $H^{2}=8\pi G\Lambda $
and $r_{0}=\sqrt{\frac{Q}{b}}$. However, the above non-linear differential
equation has several integrability problems that can be avoided taken $%
2G\left( b^{2}-4\pi \Lambda \right) /3=0$ . The solution of (41) with $%
2G\left( b^{2}-4\pi \Lambda \right) /3=0$ is given by : 
\begin{equation}
\begin{array}{l}
\pm \sqrt{-B}\left( \tau -\tau _{0}\right) = \\ 
\\ 
=a^{2}\left( 12+6\left( \frac{a}{r_{0}}\right) ^{4}+\left( \frac{a}{r_{0}}%
\right) ^{8}\right) ^{1/4}F_{1}\left[ 1/2,1/4,1/4,3/2,-\frac{\left( \frac{a}{%
r_{0}}\right) ^{4}}{3+\sqrt{3}},\frac{\left( \frac{a}{r_{0}}\right) ^{4}}{-3+%
\sqrt{3}}\right]%
\end{array}%
\end{equation}
where $F_{1}$ is the Appell hypergeometric function, $B\equiv \frac{2}{3}%
GQ^{2}\sqrt{6}$ and the relation between the charge $Q$ and the Born-Infeld
parameter $b$ has been used. It is easy to note that we have the following
relation between the cosmological constant and the absolute field of the
Born-Infeld theory 
\begin{equation}
\frac{b^{2}}{4\pi }=\Lambda \hspace{1cm}\left( G\neq 0\right)
\end{equation}
The shape of the wormhole solution is given in Figure 1. We have now the
following different points of view for this result:

i) for the magnetic field one can consider a magnetic charge \textit{%
imaginary }producing it, $\sqrt{-B}$ is a \textit{real} value and the
cosmological constant is for an \textit{anti-de Sitter space-time.} For this
case, the solution is obviosly euclidean;

ii) on the other hand, if $b^{2}\sim \Lambda $ is positive $(\Lambda >0)$,
from (42) we can see that the solution corresponds to Lorenzian wormhole in
De-Sitter space-time and electric charge. This is in some meaning, a Wick
rotation or, mathematically speaking, an analytical prolongation of the
euclidean interpretation given above. It is: $\tau \rightarrow i\tau =t$, $%
Q_{M}\rightarrow iQ_{M}=Q_{E}$, $\Lambda <0\rightarrow \Lambda >0$ $\left(
ADS\rightarrow DS\right) $.

A more detailled analisys for this type of solutions will be given elsewhere
[26].

We show that the lagrangian proposed as a candidate for the non-abelian
Born-Infeld theory gives a very wide spectrum of gravitational exact
solutions, and also in the case of flat $O\left( 4\right) $ configurations
the structure of the proposed action fulfil the energy considerations and
the topological bound (10). As it was explained in the previous paragraphs,
the action proposed reduces automatically to the Yang-Mills form under
BPS-like condition. This means that the Tseytlin prescription of symmetrized
trace clearly is not the only one that takes the linear form under BPS
considerations, as was claimed in references [28]. We can see that from the
analysis of the previous sections that the lagrangian presented here is
consistent not only from the BPS point of view, but also from the first
principles: Independence of the gauge group, conservation of its structure
in all types of configurations, and remains closely (below) to the
topological bound (10) more than other proposals for a non-abelian
Born-Infeld action. In the next section we will make the supersymmetric
extension of our action, in order to complete the requirements clearly
analyzed in [28].

\section{Supersymmetric extension}

Having shown in the above sections several reasons to propose the action (3)
as a candidate for the non-abelian Born-Infeld lagrangian, we pass to
discuss this problem from the point of view of the supersymmetry. We already
saw, from the previous sections, that the very important property of our
proposed non abelian action is its absolute independence of the gauge group
and, for instance, also independent of any trace prescription (for
supersymmetric version of NBI with symmetric trace prescription see ref.
[32]). This fundamental point makes the supersymmetric extension of our
model not only possible but also simplest. From the NABI\ lagrangian (3) we
obtain after expansion of the determinant 
\begin{equation*}
S_{NBI}=\frac{b^{2}}{4\pi }\int \sqrt{-g}dx^{4}\left\{ 1-\sqrt[4]{\gamma
^{4}-\frac{\gamma ^{2}}{2}\overline{M}^{2}-\frac{\gamma }{3}\overline{M}^{3}+%
\frac{1}{8}\left( \overline{M}^{2}\right) ^{2}-\frac{1}{4}\overline{M}^{4}}%
\right\} 
\end{equation*}
we start for writing it in the form 
\begin{equation}
L_{NBI}=\frac{b^{2}}{4\pi }\overset{\infty }{\underset{n=0}{\sum }}%
q_{n}\left( \Gamma +\Delta \right) ^{n+1}
\end{equation}
where we defined 
\begin{equation}
\Gamma \equiv \gamma ^{4}-1
\end{equation}
\begin{equation}
\Delta \equiv -\frac{\gamma ^{2}}{2}\overline{M}^{2}-\frac{\gamma }{3}%
\overline{M}^{3}+\frac{1}{8}\left( \overline{M}^{2}\right) ^{2}-\frac{1}{4}%
\overline{M}^{4}
\end{equation}
and the coefficients $q_{n}$ are given by the following formula 
\begin{equation}
q_{n}=\left\{ 
\begin{array}{l}
\left( -\frac{1}{4}\right) ^{n+1}\frac{1}{\left( n+1\right) !}\overset{n}{%
\underset{k=1}{\prod }}\left( 4k-1\right) \ \ \ \hspace{1cm}\left( n>0\right)
\\ 
\\ 
-\frac{1}{4}\ \ \ \text{ \hspace{1cm}\hspace{1cm}\hspace{1cm}\hspace{1cm}%
\hspace{1cm}}\left( n=0\right)%
\end{array}
\right.
\end{equation}
$L_{NBI}$ can be rewritten as 
\begin{equation}
L_{NBI}=\frac{b^{2}}{4\pi }\overset{\infty }{\underset{n=0}{\sum }}q_{n}%
\overset{n+1}{\underset{j=0}{\sum }}\left( 
\begin{array}{c}
n+1 \\ 
j%
\end{array}
\right) \Gamma ^{j}\ \Delta ^{n+1-j}
\end{equation}
The basic ingredient for the N=1 supersymmetric extension of the non abelian
Born Infeld Lagrangian is the non-abelian chiral superfield $W_{\alpha }$%
\begin{equation}
W_{\alpha }\left( y,\theta \right) =\frac{1}{8}\overline{D}_{\overset{.}{%
\alpha }}\overline{D}^{\overset{.}{\alpha }}e^{-2V}D_{\alpha }e^{2V}
\end{equation}
Under a gauge transformation it transforms covariantly, 
\begin{equation*}
W_{\alpha }\rightarrow =e^{-2i\Lambda }\ W_{\alpha }\ e^{2i\Lambda } 
\end{equation*}
Its hermitian conjugate transform as 
\begin{equation*}
\overline{W}_{\overset{.}{\alpha }}\rightarrow =e^{-2i\Lambda ^{\dagger }}\ 
\overline{W}_{\overset{.}{\alpha }}\ e^{2i\Lambda ^{\dagger }} 
\end{equation*}
Written in components $W_{\alpha }$ reads 
\begin{equation*}
W_{\alpha }\left( y,\theta \right) =i\lambda _{\alpha }-\theta _{\alpha }D-%
\frac{i}{2}\left( \theta \sigma ^{\mu }\overline{\sigma }^{\nu }\right)
_{\alpha }F_{\mu \nu }-\theta \theta \left( \not{\nabla}\overline{\lambda }%
\right) _{\alpha } 
\end{equation*}
where a chiral variable was introduced 
\begin{equation*}
y^{\mu }=x^{\mu }+i\theta \sigma ^{\mu }\overline{\theta } 
\end{equation*}
with 
\begin{equation*}
F_{\mu \nu }=\partial _{\mu }A_{\nu }-\partial _{\nu }A_{\mu }+i\left[
A_{\mu },A_{\nu }\right] 
\end{equation*}
and 
\begin{equation*}
\not{\nabla}\overline{\lambda }=\left( \sigma ^{\mu }\right) _{\alpha 
\overset{.}{\alpha }}\left( \partial _{\mu }\overline{\lambda }^{\overset{.}{%
\alpha }}+i\left[ A_{\mu },\overline{\lambda }^{\overset{.}{\alpha }}\right]
\right) 
\end{equation*}
Similarly as the SUSY extension of N=1 Yang-Mills theory can be constructed
from $W^{2}$ and its hermitian conjugate $\overline{W}^{2}$, we can
construct the SUSY extension of N=1 NABI\ theory considering 
\begin{eqnarray}
\left( W^{\alpha }W_{\alpha }\right) _{\mu \nu } &=&-\frac{\eta _{\mu \nu }}{%
4}\left[ \lambda ^{2}+i\left( \theta \lambda D+D\theta \lambda \right)
-\theta \theta \left( i\lambda \not{\nabla}\overline{\lambda }+i\overline{%
\lambda }\overline{\not{\nabla}}\lambda -D^{2}\right) \right] + \\
&&+\frac{1}{2}\left[ \theta \left( \sigma _{\mu }\overline{\sigma }^{\rho
}\right) \lambda F_{\nu \rho }+F_{\mu \rho }\theta \left( \sigma _{v}%
\overline{\sigma }^{\rho }\right) \lambda -\theta \theta \left( F_{\mu \rho
}^{a}F_{\nu a}^{\ \rho }+i\widetilde{F}_{\mu \rho }^{a}F_{\nu a}^{\ \rho
}\right) \right]  \notag
\end{eqnarray}
and analogically its hermitian conjugate $\left( \overline{W}_{\overset{.}{%
\alpha }}\overline{W}^{\overset{.}{\alpha }}\right) _{\mu \nu }$ where 
\begin{equation*}
\widetilde{F}_{\mu \rho }^{a}\equiv \frac{1}{2}\varepsilon _{\mu \nu \alpha
\beta }F^{\alpha \beta a} 
\end{equation*}
Now we can define the following object 
\begin{equation}
M_{\mu \nu }\equiv -\left[ \left( W^{\alpha }W_{\alpha }\right) _{\mu \nu
}+\left( \overline{W}_{\overset{.}{\alpha }}\overline{W}^{\overset{.}{\alpha 
}}\right) _{\mu \nu }\right]
\end{equation}
with an on-shell purely bosonic part $:\left. \int \left( d^{2}\theta +d^{2}%
\overline{\theta }\right) M_{\mu \nu }\right| _{bos}=F_{\mu \rho }^{a}F_{\nu
a}^{\ \rho }.$

Now, in order to construct higher powers of $M_{\mu }^{\nu }$ $\left( 
\symbol{126}F^{2}\right) $ necessary to obtain the non-abelian lagrangian we
define the superfield 
\begin{equation}
X_{\mu \nu }\equiv \frac{1}{4}\left[ e^{2V}\overline{D}^{2}\left(
e^{-2V}\left( \overline{W}_{\overset{.}{\alpha }}\overline{W}^{\overset{.}{%
\alpha }}\right) _{\mu \nu }e^{2V}\right) e^{-2V}+e^{-2V}D^{2}\left(
e^{2V}\left( W^{\alpha }W_{\alpha }\right) _{\mu \nu }e^{-2V}\right) e^{2V}%
\right]
\end{equation}
their $\theta =0$ component give, as in the abelian BI\ case, the obvious
result 
\begin{equation*}
\left. X_{\mu }^{\mu }\right| _{\theta =0}=F_{\mu \rho }^{a}F_{a}^{\ \mu
\rho } 
\end{equation*}
and this field also transform as under generalized gauge transformations $%
X_{\mu \nu }\rightarrow =e^{-2i\Lambda }\ X_{\mu \nu }e^{2i\Lambda }.$ With
the supersymmetric gauge invariant objects $M_{\mu \nu }$ and $X_{\mu \nu },$
the corresponding traceless objects $\overline{M}_{\mu \nu }$ , $\overline{X}%
_{\mu \nu }$ and also $\gamma $, can be easily constructed: $\overline{M}%
_{\mu \nu }\equiv M_{\mu \nu }-\frac{\eta _{\mu \nu }}{4}M_{\rho }^{\rho }$
, $\overline{X}_{\mu \nu }\equiv X_{\mu \nu }-\frac{\eta _{\mu \nu }}{4}%
X_{\rho }^{\rho }$ , $\gamma =1+\frac{X_{\rho }^{\rho }}{4}$

From above considerations, we propose the following supersymmetric
Lagrangian for the non-abelian generalization of the Born-Infeld theory 
\begin{equation}
L_{NBI}=\overset{\infty }{\underset{n=0}{\sum }}C_{rst}\int \left(
d^{2}\theta +d^{2}\overline{\theta }\right) \left( \Gamma _{2}+\Delta
_{2}\right) ^{r}\ \Gamma _{0}^{s}\text{ }\Delta _{0}^{t}
\end{equation}
where 
\begin{eqnarray*}
\Delta _{2} &\equiv &-\frac{1}{2}\left[ \overline{M}_{\nu }^{\mu }\overline{X%
}_{\mu }^{\nu }\left( 1+\frac{X_{\rho }^{\rho }}{4}\right) ^{2}\right] -%
\frac{1}{3}\left[ \overline{M}_{\rho }^{\mu }\overline{X}_{\sigma }^{\rho }%
\overline{X}_{\mu }^{\sigma }\left( 1+\frac{X_{\rho }^{\rho }}{4}\right) %
\right] + \\
&&+\frac{1}{8}\left[ \left( \overline{M}_{\rho }^{\mu }\overline{X}_{\mu
}^{\rho }\right) \left( \overline{X}_{\rho }^{\mu }\overline{X}_{\mu }^{\rho
}\right) \right] -\frac{1}{4}\left( \overline{M}_{\mu }^{\nu }\overline{X}%
_{\lambda }^{\mu }\overline{X}_{\rho }^{\lambda }\overline{X}_{\nu }^{\rho
}\right)
\end{eqnarray*}
\begin{eqnarray*}
\Delta _{0} &\equiv &-\frac{1}{2}\left[ \overline{X}_{\nu }^{\mu }\overline{X%
}_{\mu }^{\nu }\left( 1+\frac{X_{\rho }^{\rho }}{4}\right) ^{2}\right] -%
\frac{1}{3}\left[ \overline{X}_{\rho }^{\mu }\overline{X}_{\sigma }^{\rho }%
\overline{X}_{\mu }^{\sigma }\left( 1+\frac{X_{\rho }^{\rho }}{4}\right) %
\right] + \\
&&+\frac{1}{8}\left[ \left( \overline{X}_{\rho }^{\mu }\overline{X}_{\mu
}^{\rho }\right) ^{2}\right] -\frac{1}{4}\left( \overline{X}_{\mu }^{\nu }%
\overline{X}_{\lambda }^{\mu }\overline{X}_{\rho }^{\lambda }\overline{X}%
_{\nu }^{\rho }\right)
\end{eqnarray*}
\begin{equation*}
\Gamma _{2}\equiv M_{\rho }^{\rho }\left[ 1+\frac{3}{8}\left( X_{\rho
}^{\rho }\right) +\frac{1}{16}\left( X_{\rho }^{\rho }\right) ^{2}+\frac{1}{%
256}\left( X_{\rho }^{\rho }\right) ^{3}\right] 
\end{equation*}
\begin{equation*}
\Gamma _{0}\equiv X_{\rho }^{\rho }\left[ 1+\frac{3}{8}\left( X_{\rho
}^{\rho }\right) +\frac{1}{16}\left( X_{\rho }^{\rho }\right) ^{2}+\frac{1}{%
256}\left( X_{\rho }^{\rho }\right) ^{3}\right] 
\end{equation*}
Notice that it remains to determine the arbitrary coefficients $C_{rst}$
imposing the condition that the bosonic sector of the theory does coincide
with the NABI lagrangian. The particular choice for the coefficients $%
C_{rst} $ that leads the purely bosonic part of NABI\ lagrangian (3) is the
following 
\begin{equation*}
C_{0,s,t}=0\text{ \ \ for all }s,t 
\end{equation*}
\begin{equation}
C_{1,j,n-j}=\left\{ 
\begin{array}{l}
\left( \frac{b^{2}}{4\pi }\right) q_{n}\left( 
\begin{array}{l}
n \\ 
j%
\end{array}
\right) \ \ \ \text{for }j\leq n \\ 
\\ 
0\ \ \ \ \text{ \hspace{1cm}\hspace{1cm}\ for }j>n%
\end{array}
\right. \ \text{ }
\end{equation}
With the knowledge of the coefficients $C_{1,j,n-j}$ the supersymmetric
lagrangian can be easily written in the form$\left. L_{NBI}\right|
_{susy}=L_{NBI}+L_{fer}+L_{fb}$ where $L_{NBI}$ is the purely bosonic
lagrangian (putting fermions to zero), $L_{fb}$ includes kinetic fermion and
crossed boson-fermion terms and $L_{fer}$ contains self-interacting fermion
terms.

We have then been able to construct a $N=1$ supersymmetric non-abelian
lagrangian (53) in the context of superfields formulation, with a bosonic
part expressed in terms of the fourth root of $\left| g\right| \ det\left(
g_{\mu \nu }+\mathcal{F}_{\ \mu \lambda }^{a}\mathcal{F}_{a\nu }^{\ \ \
\lambda }\right) $. As usual, we have employed the natural curvature
invariants as building blocks in the superfield construction arriving to a
lagrangian which, in its bosonic sector, depends only on the invariants
given by expressions (3b), and it is expressed without any trace
prescription. Odd powers of the field strength $F$ are absent because our
starting point was the abelian equivalent fourth root version of the
Born-Infeld lagrangian (2) where the basic object of its structure is $%
M_{\mu \nu }=F_{\mu \sigma }F_{\nu }^{\sigma }$ . Notice also there exists
the technical impossibility to construct a superfield functional of $W$ and $%
DW$ ($\overline{W}$ and $\overline{D}\overline{W}$) containing $F^{3}$ in
its higher $\theta $ component.

With the supersymmetric extension of our proposed non-abelian Born-Infeld
action, and having account of all considerations and results from the
previous sections in the different contexts: Born-Infeld theory itself,
D-brane and superstring theory, we show that the action (3) is a strong
candidate towards a concrete non-abelian generalization of the Born-Infeld
theory .

However, it will be interesting to analyze the non-abelian Born-Infeld
action from the point of view of non-linear realizations [18, 19] as shown
by J. Bagger and A. Galperin in [17] for the abelian case. This will be our
task in the near future [26].

\section{Concluding remarks}

In this work a new non-abelian generalization Born-Infeld action is proposed
from a geometrical point of view. The advantage of this form of the
non-abelian Born-Infeld action with the other attempts is based in several
points:

1) the process of non-abelianization of the new action is perfomed in the
more natural form and is based on a geometrical propierty of the abelian
Born-Infeld lagrangian in its determinantal form;

2) in the new action there is no trace prescriptions;

3) the new action has full independence of the gauge group;

4) by construction, our lagrangian satisfies the Minkowski's inequality
(topological bound), saturating the bound when the (anti-)self-duality
condition (10) is inserted in (3). In this case, the NBI\ lagrangian
linearizes satisfying also the BPS-like conditions [28], becoming to the
Yang-Mills lagrangian.

5) from the point 4) we can see that our action is minimal remained closely
(below) the topological bound more than the other non-abelian lagrangians
proposed.

6) the supersymmetrization of the model is performed showing that the
proposed action fulfil the requirements given by the BPS-SUSY relations;

7) by analogy with the abelian case, the lagrangian proposed satisfies all
the following properties

i) we find the ordinary Yang-Mills theory in the limit $b\rightarrow \infty
. $

ii) the electric components $F_{\mu 0}^{a}$ non-abelian electromagnetic
tensor should be bounded for (i) when the magnetic components vanish

iii) the action is invariant under diffeomorphisms .

iv) the action is real;

8) a static spherically symmetric regular solution of the Einstein-NABI\
system for an isotopic ansatz is solved and the asymptotic behaviour of the
solution is in agreement with the type of solutions given by Ikeda and
Miyachi [9] for Yang-Mills in flat space and a new instanton-wormhole
solution in non-abelian Born-Infeld-Einstein theory is presented. We show
that the lagrangian proposed as a candidate for the non-abelian Born-Infeld
theory gives gravitational exact solutions that, in the case of the
wormhole, one can see the following:

i) there exists a link between the abolute field of Born and Infeld $b$ and
the cosmological constant $\Lambda $, and both ($\Lambda $ and $b$) can be
identified;

ii) the general shape of the wormhole and the tunnel radius are driven by
the Born-Infeld theory itself without necessity of to introduce any
additional field in particular.

This means that the NABI\ generalization presented here has a fundamental
importance and physical meaning not only theoretically, but also from the
phenomenological point of view.

In the context of a general analysis of possible NABI\ actions, the NABI
generalization presented in this work is an strong candidate to describe the
low-energy dynamics of D-branes, the solitons in the nonperturbative
spectrum of the (super) string theory. Is must be noted that, as was
suggested in ref.[37], any candidate for the full NABI action should be
reproduce the correct spectra (spacing of energy levels) in a gauge theory
configuration whose dual corresponds to 2-branes on $T^{4}$ where the
Tseytlin's symmetrized trace prescription fails. We believe that our
proposed action can be resolve or improve this situation because the good
properties given above. It is the task for a forthcoming paper [26] when we
will reproduce with our \thinspace action the explicit computations given in
[37]. .

It is interesting to note that in our work we take the antisymmetric
background tensor field $B_{\mu \nu }=0$ for simplicity, but its importance
is very well know from the point of view of the (super) string theories
where it firstly was associated with massless modes[41] and recently the
presence of such a background for the string dynamics has the very important
implication of non-commutativity of the space-time [42], and in many
supergravity models [43]. Recently a careful analysis of the $B_{\mu \nu }$
field in a five-dimensional brane-world scenario where it was associated
with torsion was made [44]. As a result of this analysis was shown that it
is possible to construct a topological configuration like a static cosmic
string on the brane whose formation involves only the zero mode. We can
expect that such type of configurations also can appear from our non abelian
generalization of the Born-Infeld action in a similar five-dimensional
scenario (i.e: Randall-Sundrum type). Also we expect that non-commutativity
of the space-time will be obtained as a result of switch-on the $B_{\mu \nu
} $ field in the proposed new NABI\ action. These points will be discussed
with more detail in the near future [26].

However, the geometry of the lagrangian density is only a piece of the whole
picture. The full perspective about the advantages of the second order
actions with the different approaches in the treatment of the non-abelian
extensions to Born-Infeld lagrangian will be given elsewhere [26], when we
put focus on D-branes actions, duality and quantization.

\section{Acknowledgements}

This work is in memory of my superviser professor Anatoly I. Pashnev, a
great man and a great scientist. I am very grateful to professors G. N.
Afanasiev, B. M. Barbashov and in particular to E. A. Ivanov for very useful
discussions and suggestions in crucial parts of this research. Thanks also
are given to the people of the Bogoliubov Laboratory of Theoretical Physics
and Directorate of the JINR\ for their hospitality and support.

\section{Appendix}

Explicity, the components of the Einstein tensor for the line element (21)
are \bigskip 
\begin{equation}
G^{0}\,_{0}=e^{-2\Phi }\Psi -e^{-2F}
\end{equation}
\begin{equation*}
\Psi \equiv \left[ 2\partial _{r}\partial _{r}F-2\partial _{r}\Phi
\,\partial _{r}F+3\left( \partial _{r}F\right) ^{2}\right] 
\end{equation*}
\begin{equation}
G^{1}\,_{1}=e^{-2\Phi }\left[ 2\partial _{r}\Lambda \,\partial _{r}F+\left(
\partial _{r}F\right) ^{2}\,\right] -e^{-2F}
\end{equation}
\begin{equation}
G^{2}\,_{2}=G^{3}\,_{3}=e^{-2\Phi }\left[ \partial _{r}\partial _{r}\left(
F+\Lambda \right) -\partial _{r}\Phi \,\partial _{r}\left( F+\Lambda \right)
+\left( \partial _{r}\Lambda \right) ^{2}+\left( \partial _{r}F\right)
^{2}+\partial _{r}F\,\partial _{r}\Lambda \right]
\end{equation}

\begin{equation}
G^{1}\,_{3}=G^{2}\,_{3}=G^{0}\,_{3}=G^{0}\,_{2}=G^{0}\,_{1}=G^{1}\,_{2}=0
\end{equation}
The components of the Born-Infeld energy-momentum tensor for SU(2) gauge
group in the tetrad (22) are 
\begin{equation}
-T_{00}=T_{11}=\frac{b^{2}}{4\pi }\left( 1-\sqrt{\left( \frac{r_{0}}{e^{G}}%
\right) ^{4}+1}\right)
\end{equation}
\begin{equation}
T_{22}=T_{33}=\frac{b^{2}}{4\pi }\left( 1-\frac{1}{\sqrt{\left( \frac{r_{0}}{%
e^{G}}\right) ^{4}+1}}\right)
\end{equation}
in the expressions (59-60)\ we utilized the isotopic ansatz and the
expression (27) for $F_{01}^{a}.$

\section{References\ \ \ \ \ \ }

[1] M. Born and L. Infeld, Proc. Roy. Soc.(London) \textbf{144}, 425 (1934).

[2] C. Misner, K. Thorne and J. A. Wheeler, \textit{Gravitation}, (Freeman,
San Francisco, 1973), p. 474.

[3] M. Born , Proc. Roy. Soc. (London) \textbf{143}, 411 (1934).

[4] T. Hagiwara , J. Phys. A: Math. and Gen.\textbf{14}, 3059 (1981).

[5] B. M. Barbashov and N. A. Chernikov, Zh. Eksp. Theor. Fiz. (JETP) 
\textbf{50}, 1296 (1966); \textbf{51}, 658 (1966); Comm. Math. Phys. \textbf{%
5}, 313 (1966).

[6] S. Deser and R. Puzalowski, J. Phys A: Math. and Gen. \textbf{13}, 2501
(1980).

[7] M. Duff and C. M. Isham,\ Nucl. Phys.\textbf{\ B} \textbf{162}, 271
(1980).

[8] J. Plebanski, \textit{Lectures in Non-Linear electrodynamics},
(NORDITA-Lecture notes, 1968).

[9] M. Ikeda and Y. Miyachi, Prog. Theor. Phys. \textbf{27}, 424 (1962).

[10] R. Roskies, Phys. Rev. \textbf{D}. \textbf{15}, 1722 (1977).

[11] A. Tseytlin, Nucl. Phys. \textbf{B 501}, 41 (1997).

[12] A. S. Prudnikov, Yu. Brychov and O. Marichev, \textit{Integrals and
Series }(Gordon and Breach, New York, 1986).

[13] D. J. Cirilo Lombardo, Journ. of . Math. Phys. \textbf{46}, 042501
(2005), D. J. Cirilo Lombardo, Preprint JINR-E2-2003-221.

[14] S. Cecotti and S. Ferrara, Phys. Lett \textbf{B} \textbf{187}, 335
(1987).

[15] M. Aboud Zeid and C. M. Hull, Phys. Lett. \textbf{B} \textbf{428}, 277
(1998).

[16] E. Serie' et al., Phys. Rev. D \textbf{68}, 125003 (2003).

[17] J. Bagger and A. Galperin, Phys. Rev. \textbf{D} \textbf{55}, 1091
(1997).

[18] E. A. Ivanov and V. I. Ogievetsky, Theor. Math. Phys. \textbf{25},1050
(1975); Lett. Math. Phys.\textbf{\ 1}, 309 (1976).

[19] E. A. Ivanov and B. M. Zupnik, Nucl. Phys. \textbf{B} \textbf{318}, 3
(2001).

[20] V. Tapia: arXiv: math-ph/0208010

[21] I. M. Gelfand, M. M. Kapranov and A. V. Zelevinsky, Adv. Math. \textbf{%
96}, 226 (1992).

[22] A. Cayley, Cambridge Math. J. \textbf{4}, 193 (1845).

[23] P. A. M. Dirac, Rev Mod. Phys. \textbf{34}, 592 (1962).

[24] M. Aboud Zeid,\textit{\ Actions for curved branes}, hep-th/0001127.

[25] D. Brecher and M. J. Perry, Nucl. Phys.\textbf{\ B} \textbf{566}, 151
(2000).

[26] D. J. Cirilo-Lombardo, in preparation.

[27] G. W. Gibbons and K. Hashimoto, JHEP \textbf{9}, 13 (2000).

[28] D. Brecher, \textit{BPS states of NBI action, }hep-th/9804180.

[29] J. H. Park, Phys. Lett. \textbf{B} \textbf{458}, 471 (1999).

[30] A. Hosoya and W. Ogura, Phys. Lett. \textbf{B} \textbf{225}, 117 (1989).

[31] S. B. Giddings and A. Strominger, Nucl. Phys. \textbf{B} \textbf{306},
890 (1988).

[32] S. Gonorazky et al., Phys. Lett. \textbf{B} \textbf{449}, 187 (1999).

[33] A. Belavin et al., Phys. Lett. \textbf{B} \textbf{59}, 85 (1975).

[34] G. W. Gibbons M. B. Green and M. J. Perry, Phys. Lett. \textbf{B} 
\textbf{370}, 37 (1996).

[35] C.S Chu, P. M. Ho and Y. Y. Wu, Nucl. Phys. \textbf{B} \textbf{541},
179 (1999).

[36] S. R. Das, S. K. Rama and S. P. Trivedi, JHEP \textbf{3}, 4 (2000).

[37] A. Hashimoto and W. Taylor IV, Nucl. Phys.\textbf{\ B 503,} 193 (1997).

[38] R. G. Leigh, Mod. Phys. Lett. \textbf{A 4}, 2767, 2073 (1989).

[39] G. W. Gibbons, Nucl Phys. \textbf{B 514}, 603 (1998).

[40] E. Witten, Nucl Phys. \textbf{B 460}, 335 (1996).

[41] M. Kalb and P. Ramond, Phys. Rev. \textbf{D 9}, 2273 (1974).

[42] N. Seiberg and E. Witten, JHEP \textbf{9}, 32 (1999).

[43] S. J. Gates, M. Grisaru, M. Rocek and W. Siegel, \textit{Superspace}
(W. A. Benjamin, New York, 1983).

[44] N. R. F. Braga and C. N. Ferreira, JHEP \textbf{3}, 39 (2005).

\bigskip

\bigskip

\bigskip

\bigskip

\bigskip

\bigskip

\bigskip

\bigskip

\bigskip

\bigskip

\end{document}